\documentclass[11pt]{article}
\usepackage{amssymb}
\usepackage{amsmath}
\usepackage{graphics}
\usepackage{epsfig}
\usepackage{a4wide}
\usepackage{cite}
\usepackage{multirow}
\usepackage{color}
\usepackage{mathrsfs}
\usepackage{subfigure}
\usepackage{hep}
\usepackage{diagbox}

\DeclareGraphicsExtensions{.eps}
\begin{document}

\title{ $WW\gamma$ production at hadron colliders with NLO QCD+EW corrections and parton shower effects }
\author{
Jian-Wen Zhu$^{1,2}$, Ren-You Zhang$^{1,2}$\footnote{Corresponding author. zhangry@ustc.edu.cn}, Wen-Gan Ma$^{1,2}$, Qiang Yang$^{1,2}$, and Yi Jiang$^{1,2}$ \\ \\
{\small $^1$ State Key Laboratory of Particle Detection and Electronics,} \\
{\small University of Science and Technology of China, Hefei 230026, Anhui, People's Republic of China} \\
{\small $^2$ Department of Modern Physics, University of Science and Technology of China,}  \\
{\small Hefei 230026, Anhui, People's Republic of China}
}

\date{}
\maketitle
\vskip 10mm

\begin{abstract}
$W^+W^-\gamma$ production in proton-proton collision provides a window to the mechanism of electroweak symmetry breaking and a direct accessment to triple and quartic gauge couplings. Precision study of gauge boson self-interactions may also provide evidence of existence of new physics beyond the Standard Model. In this paper, we study the $W^+W^-\gamma$ production at the LHC and future higher energy proton-proton colliders at the QCD+EW NLO including parton shower effects. We find that the contributions from the photon-induced (i.e., $q\gamma$- and $\gamma\gamma$-initiated) channels are non-negligible since the photon luminosity can be enhanced significantly with the increment of colliding energy, and the large real jet emission QCD and EW corrections can be depressed sufficiently by applying the jet veto event selection scheme. Moreover, we also investigate the theoretical errors arising from the PDF uncertainty and the factorization/renormalization scale dependence.
\end{abstract}


\vfill \eject
\baselineskip = 0.32in
\makeatletter
\@addtoreset{equation}{section}
\makeatother
\vskip 5mm
\renewcommand{\theequation}{\arabic{section}.\arabic{equation}}
\renewcommand{\thesection}{\Roman{section}.}
\newcommand{\nb}{\nonumber}

\section{Introduction}
\par
After the discovery of $125~ {\rm GeV}$ Standard Model (SM) Higgs boson \cite{ATLAS-Higgs,CMS-Higgs}, exploring the existence of new physics beyond the SM has become one of the most significant tasks of the CERN Large Hadron Collider (LHC). Therefore, both theoretical predictions and experimental measurements with higher precision are indispensable for LHC Run II and future proton-proton colliders. The non-Abelian $SU(2)_L \otimes U(1)_Y$ gauge symmetry in the SM predicts the existence of gauge boson self-interactions. The direct investigation of gauge boson self-interactions provides a crucial test of gauge structure of the SM. Multiple gauge boson production at high energy colliders provides an opportunity for precision measurement of triple and quartic gauge boson couplings, and it would help us to better understand the electroweak symmetry breaking \cite{tgc,qgc1,qgc2}. So far the gauge boson pair productions have been experimentally measured at the LHC \cite{diboson1,diboson2,diboson3,diboson4,diboson5}, and one found there is no significant deviation from the SM prediction. Moreover, in order to study the quartic gauge couplings, experimental measurements for triple gauge boson productions at the LHC, such as $pp \rightarrow WW\gamma + X$, $pp \rightarrow WZ\gamma + X$ and $pp \rightarrow WWW + X$, have drawn attention in recent few years \cite{WVA-ex,WWW-ex}.

\par
The $W^+W^-\gamma$ production has an advantage in probing sensitively both triple and quartic gauge boson self-interactions, particularly the $WWZ\gamma$ and $WW\gamma\gamma$ gauge couplings. The theoretical predictions for $W^+W^-\gamma$ production with subsequent $W$-boson decay at the LHC have been calculated up to the QCD next-to-leading order (NLO) \cite{WWA-QCD}. The electroweak (EW) corrections to $W^+W^-\gamma$ production at the ILC have been investigated in Ref.\cite{cc-WWA}, however, the EW corrected predictions at hadron colliders are still missing. The $W^+W^-\gamma$ production is a considerable SM background to associated $H \gamma$ production \cite{Hgamma} with subsequent decay $H \rightarrow W^+W^-$, where $H$ represents an exotic neutral Higgs boson of new physics beyond the SM. Moreover, the $W^+W^-\gamma$ production is also an irreducible SM background for search for singly charged Higgs boson in associated $H^{\pm}W^{\mp}$ production with subsequent decay $H^{\pm} \rightarrow W^{\pm} \gamma$ \cite{Wgamma1,Wgamma2}. The NLO QCD+EW corrections to some other triple gauge boson production processes have been widely studied \cite{WWZ,WZZ,WWW1,WWW2,WWW3,ZZA,AAV,ZZZ}. In this paper, we investigate in detail the $W^+W^-\gamma$ production at proton-proton colliders in the SM at the QCD+EW NLO including parton shower (PS) effects. The rest of this paper is organized as follows: In Section II, we present the calculation strategies for $W^+W^-\gamma$ production at the QCD+EW NLO, including the electric charge renormalization scheme, the technique for infrared (IR) singularity separation and the photon isolation criterion. In Section III, we provide the numerical results of the integrated cross sections at some typical colliding energies and some kinematic distributions of final particles at the $14~ {\rm TeV}$ LHC, and discuss in detail the theoretical uncertainties from the parton distribution functions (PDFs) and factorization/renormalization scale. Finally, we give a short summary at Section IV.

\vskip 10mm
\section{Calculation strategy}
\subsection{General setup and LO calculation}
\par
At the tree level, the $WW\gamma$ events can be produced via quark-antiquark and photon-photon annihilation channels at a proton-proton collider, i.e.,
\begin{eqnarray}
\label{channel-1}
pp \rightarrow q\bar{q} \rightarrow W^+W^-\gamma + X
~~~~~~
(q = u,~ d,~ s,~ c,~ b),
\end{eqnarray}
and
\begin{eqnarray}
\label{channel-2}
pp \rightarrow \gamma\gamma \rightarrow W^+W^-\gamma + X.
\end{eqnarray}
The light-quarks $u$, $d$, $s$, $c$ and $b$ are treated as massless particles, thus our calculation will encounter IR divergence in some specific phase-space regions where the final-state photon is soft enough, or collinear to one of initial-state massless quarks. In order to avoid IR divergence in leading order (LO) calculation, we apply the following transverse momentum and pseudorapidity cuts on the final-state photon:
\begin{eqnarray}
\label{basic cuts}
p_T^{\gamma} > 20~ {\rm GeV},
~~~~~~~~
\left| \eta^{\gamma} \right| < 2.5.
\end{eqnarray}
In both LO and NLO QCD+EW calculations, we adopt the 't Hooft-Feynman gauge. We use {\sc FeynArts-3.7} \cite{feynart} to generate all the LO and NLO QCD and EW Feynman diagrams and the corresponding amplitudes for partonic processes, and employ {\sc FormCalc-7.3} \cite{formcalc} to implement amplitude reduction and phase space integration. The scalar and tensor integrals are calculated by using {\sc LoopTools-2.8} \cite{looptools}. The QCD PS effects on the NLO QCD+EW corrected cross section and differential distributions of final $W^{\pm}$-bosons are accomplished by using above packages in combination with {\sc Pythia8} \cite{pythia}, which can be performed automatically in {\sc MadGraph5} \cite{mg5}. After the matching of NLO QCD calculation to parton shower, we obtained HepMC event file and then analyze the events by adopting {\sc MadAnalysis5} \cite{ma5}.

\par
\begin{figure}[htbp]
\centering
\includegraphics[scale=1]{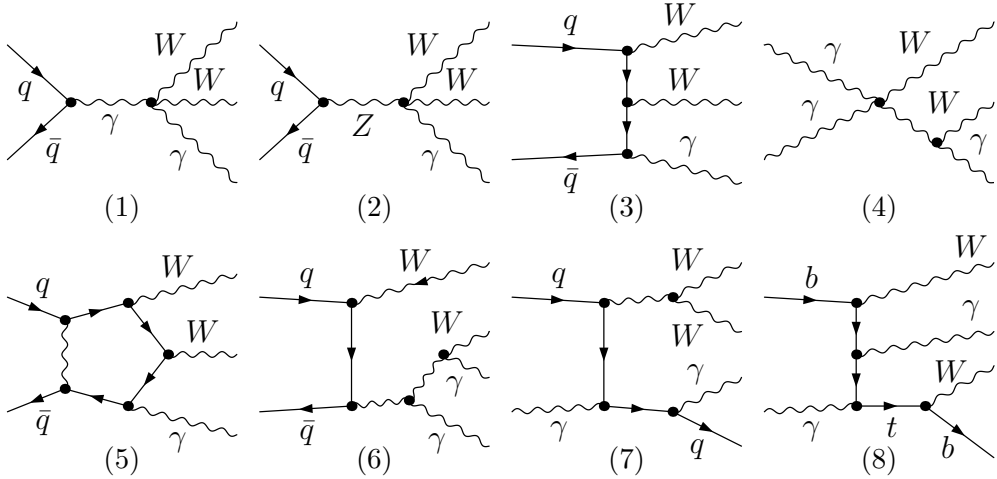}
\caption{Representative Feynman diagrams for partonic processes of $W^+W^-\gamma$ production at a proton-proton collider.}
\label{Diagrams}
\end{figure}

We depict some representative tree-level Feynman diagrams for $q\bar{q} \rightarrow W^+W^-\gamma$ in Figs.\ref{Diagrams}(1-3). It clearly shows that the $WW\gamma\gamma$ and $WWZ\gamma$ quartic gauge couplings are involved at the LO, thus the $W^+W^-\gamma$ production is an ideal channel to test these quartic gauge couplings. The NLO QCD and NLO EW corrections to the $pp \rightarrow W^+W^-\gamma + X$ process are at the ${\cal O}(\alpha_s \alpha^3)$ and ${\cal O}(\alpha^4)$, respectively. Some of the Feynman diagrams for the NLO virtual and real emission EW corrections are shown in Figs.\ref{Diagrams}(4-8). Since the photon density is normally much smaller than the densities of QCD partons ($u,~ d,~ c,~ s,~ b$ and $g$) in an energetic proton, we may ignore the high order corrections to the $\gamma\gamma$ fusion channel in our consideration. In this work, we include the LO contribution from the $\gamma\gamma$ fusion, see Fig.\ref{Diagrams}(4), as a part of EW correction to the $pp \rightarrow W^+W^-\gamma + X$ process.

\subsection{ NLO calculations}
\par
The NLO correction to the $pp \rightarrow W^+ W^- \gamma + X$ process involves the following three components: (1) virtual correction, (2) real emission correction, and (3) PDF counterterm contribution. In our calculation, both ultraviolet (UV) and IR singularities are isolated by using the dimensional regularization scheme in $D = 4 -2 \epsilon$ dimensions, and the masses and wave functions of related particles are renormalized by adopting the on-mass-shell renormalization scheme \cite{Denner}. According to the Kinoshita-Lee-Nauenberg theorem \cite{KLN}, the sum of the above three components should be IR-finite.

\subsubsection{QCD calculation}
\par
The NLO QCD correction to $pp \rightarrow W^+ W^- \gamma + X$ comes from only the quark-antiquark annihilation partonic processes $q\bar{q} \rightarrow W^+ W^- \gamma~ (q=u,~ d,~ s,~ c,~ b$), while the photon-photon fusion channel $\gamma\gamma \rightarrow W^+ W^- \gamma$ does not contribute to the $W^+W^-\gamma$ production at the QCD NLO. The NLO QCD amplitudes for $q\bar{q} \rightarrow W^+ W^- \gamma$ are contributed by vertex, box and pentagon Feynman diagrams. We see that both UV and IR singularities exist in the one-loop amplitudes, and the UV divergence can be removed after performing the renormalization procedure. Since the $W^+W^-\gamma$ production is a pure EW process at the LO (i.e., the strong coupling constant is not involved in the LO Feynman amplitude), only the wave functions and masses of related colored particles need to be renormalized at the QCD NLO. The definition of these renormalization constants in the on-mass-shell scheme can be found in Ref. \cite{Denner}. To cancel the IR singularities at the QCD NLO, the real gluon bremsstrahlung and the real light-quark emission should be considered. We adopt the Frixione-Kunsz-Signer (FKS) subtraction scheme \cite{FKS1,FKS2} to subtract the IR singularities for these real emission processes. The real gluon bremsstrahlung includes both soft and collinear IR singularities, while the real light-quark emission contains only collinear IR singularity. The soft IR singularity in the real gluon bremsstrahlung cancels exactly that in the QCD virtual correction. The collinear IR singularity in the QCD virtual correction is only partially canceled by those in the real gluon and light-quark emissions, and the remaining collinear IR singularity is absorbed by the related PDF QCD counterterms. The explicit expressions for the PDF counterterms can be found in Ref.\cite{TCPSS}.

\subsubsection{EW calculation}
\par
The LO cross section for $pp \rightarrow q\bar{q}/\gamma\gamma \to W^+W^-\gamma$ is at the ${\cal O}(\alpha^3)$. The fine structure constant $\alpha$ can be defined by a full $e^-e^+\gamma$ coupling for on-shell external electron and positron in the Thomson limit, leading to the renormalized value $\alpha = \alpha(0)$, which is called $\alpha(0)$ scheme. However, the $\alpha(0)$ scheme is not suitable for the processes containing genuine weak couplings, for example, the $W^+W^-\gamma$ production via $q\bar{q}$ annihilation considered in this work. The NLO EW corrections to those processes in the $\alpha(0)$ scheme are sensitive to the mass-singular terms $\ln (m_f^2/\mu^2)~ (f = e,~ \mu,~ \tau,~ u,~ d,~ s,~ c,~ b)$, which originate from the renormalization of the photon wave function and EW couplings. For a process with $l$ external photons and $n$ EW couplings in the LO amplitude, the mass singularities induced by $l$ external photons can only cancel those from $l$ EW coupling counterterms. Thus, the full NLO EW correction would still contain residual mass singularities from the rest $n-l$ EW coupling counterterms. To reduce the renormalization scale uncertainty at the NLO, we should take running $\alpha$ for $n-l$ EW couplings to absorb the uncanceled large logarithms. In this paper, we adopt the mixed scheme, in which the electromagnetic couplings related to external photons and the rest EW couplings at both LO and EW NLO are taken in the $\alpha(0)$ scheme and $G_\mu$ scheme \cite{Denner,Wboson,Wgamma,Zgamma,sudakov1,sudakov2}, respectively.
The fine structure constant in the $G_\mu$ scheme is given by
\begin{eqnarray}
\alpha_{G_\mu} = \frac{\sqrt{2}G_{\mu}M_W^2(M_Z^2 - M_W^2)}{\pi M_Z^2}.
\end{eqnarray}
The electric charge renormalization constant in the $\alpha(0)$ scheme \cite{Denner} can be written as
\begin{eqnarray}
\delta Z_e^{\alpha(0)}
=
-\frac{1}{2}\delta Z_{AA} - \frac{1}{2} \tan \theta_W \delta Z_{ZA}
=
\left[
\frac{1}{2}\frac{\partial \Sigma_T^{AA}(p^2)}{\partial p^2} - \tan \theta_W \frac{\Sigma_T^{AZ}(p^2)}{M_Z^2}
\right]_{p^2 = 0},
\end{eqnarray}
where $\theta_W$ is Weinberg weak mixing angle, $\delta Z_{AA}$ and $\delta Z_{ZA}$ are the wave-function renormalization constants for $\gamma-\gamma$ and $\gamma-Z$ transitions, and $\Sigma_T^{AA}$ and $\Sigma_T^{AZ}$ are the unrenormalized transverse self-energies. In the $G_{\mu}$ scheme, the electric charge renormalization constant should be modified as
\begin{eqnarray}
\delta Z_e^{G_{\mu}} = \delta Z_e^{\alpha(0)} - \frac{1}{2}\Delta r,
\end{eqnarray}
where the subtraction term $\Delta r$ is given by the one-loop EW correction to the muon decay \cite{muon-decay}.

\par
In the calculation of one-loop virtual corrections, all the 5-point (scalar and tensor) integrals induced by pentagon Feynman diagrams, e.g., Fig.\ref{Diagrams}(5), are reduced to 4-point integrals by the method proposed by Denner and Dittmaier \cite{reduce}, and the $n$-point tensor integrals $(n \le 4)$ are reduced to scalar integrals recursively by Passarino-Veltman algorithm \cite{PV}. The reduction of tensor integrals and the numerical calculation of scalar integrals are performed by using {\sc LoopTools-2.8} package. As we know, the Passarino-Veltman reduction would induce numerical instability at some phase-space region with small Gram determinant in loop calculation. This instability problem is coped with by our developed codes, which are based on {\sc LoopTools-2.8} and can switch to the quadruple precision in the instability region automatically \cite{WWW1}.

\par
For the NLO EW corrections, the IR singularities in the real photon and jet emissions are isolated by adopting the two cutoff phase space slicing method \cite{TCPSS}, which is intuitive and simple to implement. In this method, two cutoffs, $\delta_s$ and $\delta_c$, are introduced to decompose the phase space into three regions: soft region, hard collinear region and hard noncollinear region. The integrated cross sections over the soft and hard collinear regions are calculated analytically by using the soft and collinear approximations, respectively, while the integration over the hard noncollinear region is performed numerically by adopting the Monte Carlo technique. As we expect, the sum of the cross sections over these phase-space regions is independent of the soft cutoff $\delta_s$ in the range of $\delta_s \in [10^{-3},~ 10^{-5}]$ with $\delta_c = \delta_s / 50$.

\subsubsection{Event identification and selection}
\par
The real emission processes listed in Table \ref{Real-emission} may contribute to the $W^+W^-\gamma$ production at the QCD and EW NLO. The final state of each real emission process contains two massless particles (photon and jet). We define the separation of two massless tracks ``1'' and ``2'' as
\begin{eqnarray}
R_{12} = \sqrt{(\eta_1 - \eta_2)^2 + (\phi_1 - \phi_2)^2},
\end{eqnarray}
where $\eta_i$ and $\phi_i$ $(i=1,~ 2)$ are the pseudorapidities and azimuthal angles, respectively. For the real photon bremsstrahlung $q\bar{q} \rightarrow W^+W^-\gamma \gamma$, the two photon tracks in the final state are clustered into one quasi-photon and the final state is regarded as a $W^+W^-\gamma$ event, if the two photon tracks are sufficiently collinear, i.e., $R_{\gamma\gamma} \leqslant 0.1$\footnote{If $R_{\gamma\gamma} > 0.1$, the final state is regarded as a $W^+W^-\gamma\gamma$ event.}. For the real jet emission processes $q\bar{q} \rightarrow W^+W^-\gamma g$, $qg \rightarrow W^+W^-\gamma q$ and $q\gamma \rightarrow W^+W^-\gamma q$, the final-state photon and jet tracks are recombined into one quasi-particle if they are not well separated, i.e., $R_{\gamma j} \leqslant R_0$\footnote{If $R_{\gamma j} > R_0$, the final state is identified as a $W^+W^-\gamma + jet$ event.}, where $R_0$ is taken to be $0.5$ as a threshold to decide whether the photon and jet tracks can be separated unambiguously. Once the collinear photon-jet system is recombined, the final state is treated as a $W^+W^-\gamma$ event if the energy fraction of photon inside the photon-jet system exceeds a certain threshold, i.e., $z_{\gamma} \equiv \dfrac{E_{\gamma}}{E_{\gamma} + E_j} \geqslant z_{\gamma}^{{\rm cut}}$, where $z_{\gamma}^{{\rm cut}}$ is typically chosen to be $0.9$; otherwise it is treated as a $W^+W^-+jet$ event and thus should be rejected. However, in this naive event identification criterion, the final-state collinear IR divergences induced by the residual jet activities in the collinear photon-jet system from the gluon- and photon-initiated light-quark emissions, $qg \rightarrow W^+W^-\gamma q$ and $q\gamma \rightarrow W^+W^-\gamma q$, can not be canceled. To solve this problem, we should modify the above event identification criterion. In this work, we adopt the Frixione method \cite{Frix} to isolate the $W^+W^-\gamma$ events for real jet emission channels. In the Frixione isolation method, a collinear photon-jet system (i.e., $R_{\gamma j} \leqslant R_0$) is clustered into a quasi-photon only if
\begin{eqnarray}
\dfrac{p_{T}^{j}}{p_{T}^{\gamma}} \leqslant \chi(R_{\gamma j}),
\end{eqnarray}
where the restriction function $\chi(R_{\gamma j})$ is given by
\begin{eqnarray}
\chi(R_{\gamma j}) = \epsilon_{\gamma}\left(\frac{1 - \cos R_{\gamma j}}{1 - \cos R_0}\right)^n,
\end{eqnarray}
and the isolation parameter $\epsilon_{\gamma}$ and the weight factor $n$ are both set to be $1$ in the numerical calculation. Since $\lim_{R_{\gamma j} \rightarrow 0} \chi(R_{\gamma j}) = 0$, the Frixione isolation criterion retains the soft jet activity but forbids any hard collinear jet activity in the collinear photon-jet system when this collinear photon-jet system is identified as a quasi-photon. Thus, the IR singularities can also be exactly canceled after applying the event identification and selection criterion.

\par
In the five-flavor scheme, the $W^+W^-\gamma$ events produced via $\rightarrow b g/\gamma \rightarrow W^+W^-\gamma b$ and $\bar{b} g/\gamma \rightarrow W^+W^-\gamma \bar{b}$ channels are mainly from the on-shell $W^-\gamma t$ and $W^+\gamma \bar{t}$ production with subsequent top-quark decay $\overset{_{(-)}}{t} \rightarrow W^{\pm} \overset{_{(-)}}{b}$. These events are treated as $W \gamma t$ associated production, and thus should be subtracted from our calculation to avoid double counting and to keep the convergence of the perturbative description of $W^+W^-\gamma$ production. Since we assume the efficiency of $b$-tagging is $100\%$, these events can be completely excluded by applying $b$-jet veto.
\begin{table}[htbp]
\begin{center}
\renewcommand\arraystretch{1.8}
\begin{tabular}{cl|ll}
\hline
\hline
Photon bremsstrahlung & & ~~ Jet emission & \\
\hline
\multirow{3}*{$q\bar{q} \rightarrow W^+W^-\gamma\gamma$} & \multirow{3}*{$\alpha(0)\alpha_{G_\mu}^3$} &
~$q\bar{q} \rightarrow W^+W^-\gamma g$ & ~$\alpha(0) \alpha^2_{G_\mu} \alpha_s$ \\
& &
~$qg \rightarrow W^+W^-\gamma q$~ (gluon-induced) &
~$\alpha(0) \alpha^2_{G_\mu} \alpha_s$ \\
& &
~$q\gamma \rightarrow W^+W^-\gamma q$~ (photon-induced) &
~$\alpha(0) \alpha^3_{G_\mu}$ \\
\hline
\hline
\end{tabular}
\caption{Real emission channels related to the NLO QCD+EW corrections to $W^+W^-\gamma$ production at proton-proton colliders.}
\label{Real-emission}
\end{center}
\end{table}

\vskip 10mm
\section{Numerical results and discussion}
\subsection{Input parameters}
\par
The SM input parameters used in this paper are taken as \cite{PDG}:
\begin{eqnarray}
&& M_W = 80.385~{\rm GeV}, ~~ M_Z = 91.1876~{\rm GeV}, ~~ M_t = 173.21~{\rm GeV}.
\nonumber \\
&& M_H = 125.09~{\rm GeV}, ~~ G_{\mu} = 1.16638 \times 10^{-5}~{\rm GeV}^{-2}, ~~ \alpha(0) = 1 / 137.036.
~~~~~~
\end{eqnarray}
The Cabiboo-Kobayashi-Maskawa matrix is set to ${\bf 1}_{3 \times 3}$. In both LO and NLO calculations, we adopt the LUXqed\_plus\_PDF4LHC15\_nnlo\_100 ({\sc LUXqed}) PDF \cite{LUX,LUX2,PDFLHC} for initial-state parton convolution unless stated otherwise. The strong coupling constant $\alpha_s(\mu)$ is obtained by the expression in the $\overline{MS}$ scheme up to two-loop order. The factorization scale $\mu_f$ and the renormalization scale $\mu_r$ are set to be equal for simplicity, and the central scale $\mu_0$ is defined as
\begin{eqnarray}
\mu_0 = H_T / 2 = \sum_i m_{T, i} / 2,
\end{eqnarray}
where $m_{T, i} = \sqrt{m_i^2 + p_{T, i}^2}$ is the transverse mass of the final particle $i$ and the summation is taken over all the final particles.

\par
We have used both the {\sc MadGraph5} program and the {\sc FormCalc-7.3} package in LO and NLO QCD calculations to verify the correctness of our numerical calculations for the $W^+W^-\gamma$ production at the $\sqrt{s}=14~ {\rm TeV}$ LHC. With above input parameters, the LO and NLO QCD corrected integrated cross sections are obtained as
\begin{eqnarray}
&
(\text{{\sc MadGraph}})
&
~~~~~
\sigma^{{\rm LO}} = 0.18190(6)~ {\rm pb},
~~~~~
\sigma^{{\rm NLO~ QCD}} = 0.3347(5)~ {\rm pb},
\nonumber \\
&
(\text{{\sc FormCalc}})
&
~~~~~
\sigma^{{\rm LO}} = 0.18202(6)~ {\rm pb},
~~~~~
\sigma^{{\rm NLO~ QCD}} = 0.3339(5)~ {\rm pb}.
~~~~
\end{eqnarray}
We can see clearly that the numerical results obtained by using {\sc MadGraph5} and {\sc FormCalc-7.3} are in good agreement with each other within the calculation errors.

\subsection{Integrated cross sections}
\par
In order to obtain the integrated cross section including the EW corrections from the photon-induced production channels and the potentially large contribution from the interplay between the QCD and EW corrections beyond the NLO, we express the total cross section as  \cite{Wgamma}
\begin{eqnarray}
\label{NLO-exp}
\sigma^{{\rm NLO}}
=
\sigma^{{\rm LO}}
\times
\Big[
\left(1 + \delta_{q\bar{q}}^{{\rm EW}}\right)
\left(1 + \delta^{{\rm QCD}}\right)
+
\delta_{q\gamma}^{{\rm EW}}
+
\delta_{\gamma\gamma}^{{\rm EW}}
\Big],
\end{eqnarray}
where the relative corrections are defined as
\begin{eqnarray}
\label{relative}
\delta^{{\rm QCD}}
=
\frac{\Delta \sigma^{{\rm NLO~ QCD}}_{q\bar{q}} + \sigma_{qg}}{\sigma^{{\rm LO}}},
~~~
\delta_{q\bar{q}}^{{\rm EW}}
=
\frac{\Delta \sigma_{q\bar{q}}^{{\rm NLO~ EW}}}{\sigma^{{\rm LO}}},
~~~
\delta_{q\gamma}^{{\rm EW}}
=
\frac{\sigma_{q\gamma}}{\sigma^{{\rm LO}}}, ~~~
\delta_{\gamma\gamma}^{{\rm EW}}
=
\frac{\sigma_{\gamma\gamma}}{\sigma^{{\rm LO}}}.
\end{eqnarray}
The subscripts $q\bar{q},~ q\gamma$ and $\gamma\gamma$ indicate the corresponding partonic channels, and $\sigma^{{\rm LO}}$ includes only the contribution from the LO $q\bar{q}$ annihilation channel. The definition in Eq.(\ref{NLO-exp}) is preferable over the naive additive approach since it is well motivated by the large EW Sudakov logarithms at high energy scale, which leads to non-negligible high-order interplay between QCD and EW corrections \cite{mul1,mul2,mul3}.

\par
In Table \ref{CS}, we provide the numerical results of the LO and NLO QCD+EW corrected integrated cross sections, as well as the relative corrections defined in Eq.(\ref{relative}), for $W^+W^-\gamma$ production at current LHC and future proton-proton colliders in the inclusive event selection scheme\footnote{In the inclusive event selection scheme, only the baseline cuts in Eq.(\ref{basic cuts}) are imposed on the final state.}. We see that the NLO correction is dominated by the QCD contribution. The QCD $K$-factor is about $1.8$ at the $13$ and $14~ {\rm TeV}$ LHC, and increases rapidly with the increment of colliding energy and even reaches about $2.6$ at a $100~ {\rm TeV}$ proton-proton collider. Our calculation shows that the gluon-induced real jet emission channel contributes the most part of the QCD correction due to the higher luminosity of gluon in proton with increasing colliding energy. In the situation of very large NLO QCD correction, the higher order QCD contributions to the $pp \rightarrow W^+W^-\gamma + X$ process should be taken into account in precision calculation. However, the NLO QCD correction can be heavily suppressed, and thus the convergence of the perturbative QCD description can be improved, by adopting the exclusive event selection scheme, i.e., imposing a jet veto on the final-state jet. We will discuss that later in this paper.
\begin{table}[htbp]
\begin{center}
\renewcommand\arraystretch{1.8}
\begin{tabular}{ccccccc}
\hline
\hline
$\sqrt{s}$ [TeV] & $\sigma^{{\rm LO}}$ [pb] & $\sigma^{{\rm NLO}}$ [pb] & $\delta^{{\rm QCD}}$ [\%] & $\delta_{q\bar{q}}^{{\rm EW}}$ [\%]
& $\delta_{q\gamma}^{{\rm EW}}$ [\%] & $\delta_{\gamma\gamma}^{{\rm EW}}$ [\%] \\
\hline
$7$  & $0.06710(2)$ & $0.1074(1)$ & $61.5$ & $-4.10$ & $2.67$ & $2.57$ \\
$8$  & $0.08264(3)$ & $0.1356(1)$ & $65.5$ & $-4.31$ & $3.01$ & $2.76$ \\
$13$ & $0.16491(5)$ & $0.2971(3)$ & $81.3$ & $-5.00$ & $4.54$ & $3.43$ \\
$14$ & $0.18190(5)$ & $0.3327(4)$ & $84.0$ & $-5.12$ & $4.81$ & $3.54$ \\
$100$ & $1.6231(5)$ & $4.237(4)$ & $158.2$ & $-6.56$ & $14.1$ & $5.63$ \\
\hline
\hline
\end{tabular}
\caption{
LO and NLO QCD+EW corrected integrated cross sections, as well as NLO QCD and EW relative corrections, for $pp \rightarrow W^+W^-\gamma + X$ at proton-proton colliders in the inclusive event selection scheme.}
\label{CS}
\end{center}
\end{table}

\par
The contributions from the $q\gamma$- and $\gamma\gamma$-initiated channels are positive, while the NLO EW correction from the $q\bar{q}$ annihilation channel is negative and not very sensitive to the proton-proton colliding energy. At the $7$ and $8~ {\rm TeV}$ LHC, the photon-induced relative correction, given by $\delta_{\gamma\text{-}{\rm induced}}^{{\rm EW}} = \delta_{q\gamma}^{{\rm EW}} + \delta_{\gamma\gamma}^{{\rm EW}}$, compensates the NLO EW correction from the $q\bar{q}$ annihilation channel, and the full EW relative correction, defined as $\delta^{{\rm EW}} = \delta_{q\bar{q}}^{{\rm EW}} + \delta_{\gamma\text{-}{\rm induced}}^{{\rm EW}}$, is only about $1\%$. At the $13 \sim 14~ {\rm TeV}$ LHC, the full EW relative correction reaches about $3\%$ due to the increment of the photon-induced relative correction. As the increment of the $pp$ colliding energy, the $q\gamma$-initiated relative correction increases significantly, while the $q\bar{q}$- and $\gamma\gamma$-initiated relative corrections are relatively stable, especially at very high energy region. Thus, the strong cancelation between $\delta_{q\bar{q}}^{{\rm EW}}$ and $\delta_{\gamma\text{-}{\rm induced}}^{{\rm EW}}$ disappears at very high colliding energies. Table \ref{CS} clearly demonstrates that the $q\gamma$-induced relative correction increases about threefold and reaches to $14.1\%$ as the $pp$ colliding energy increases from $14~ {\rm TeV}$ to $100~ {\rm TeV}$. As we know, this large $q\gamma$-induced contribution can also be depressed by adopting the jet veto scheme in the NLO calculation. According to the above discussion, we can conclude that the total NLO EW correction to the $pp \rightarrow W^+W^-\gamma + X$ process should be considered in precise experimental measurement at high energy hadron colliders.

\par
\begin{figure}[htbp]
\centering
\includegraphics[scale=0.35]{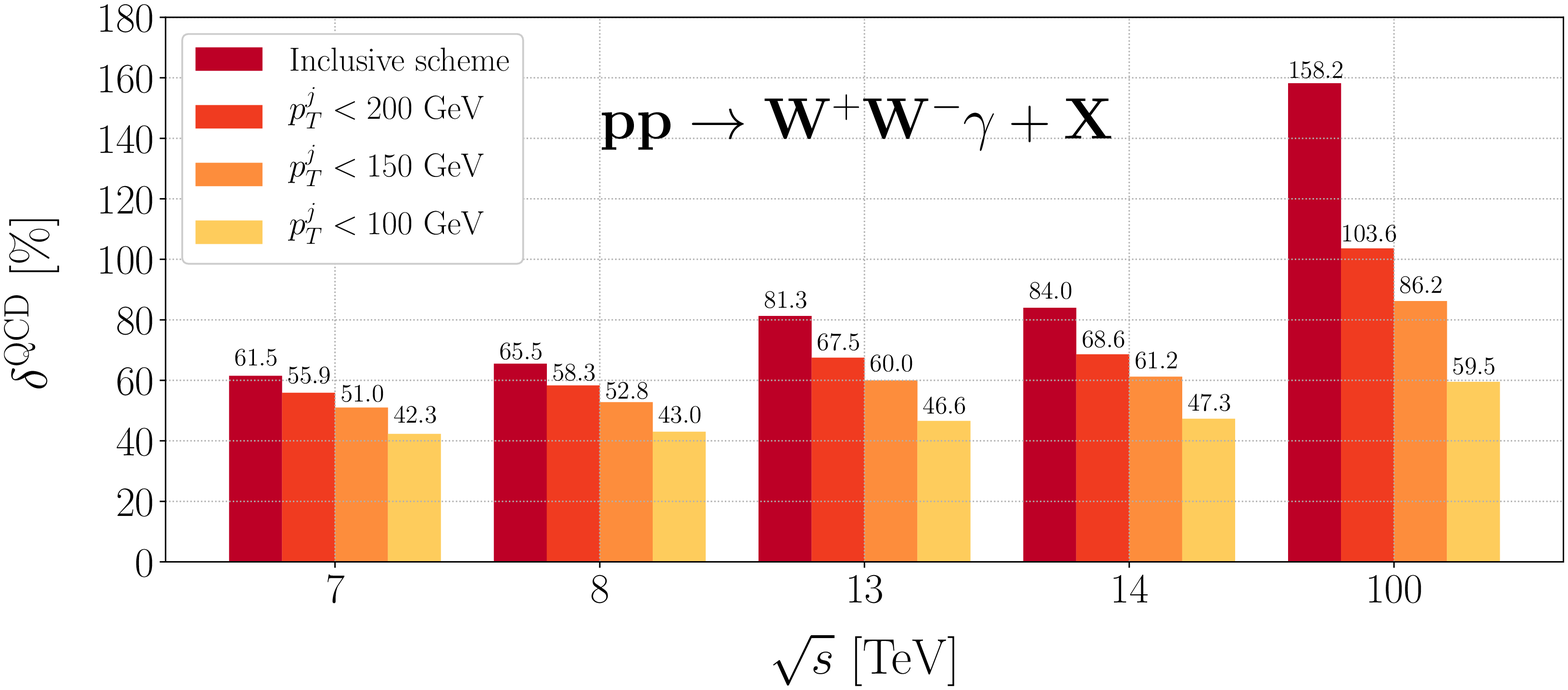}
\includegraphics[scale=0.35]{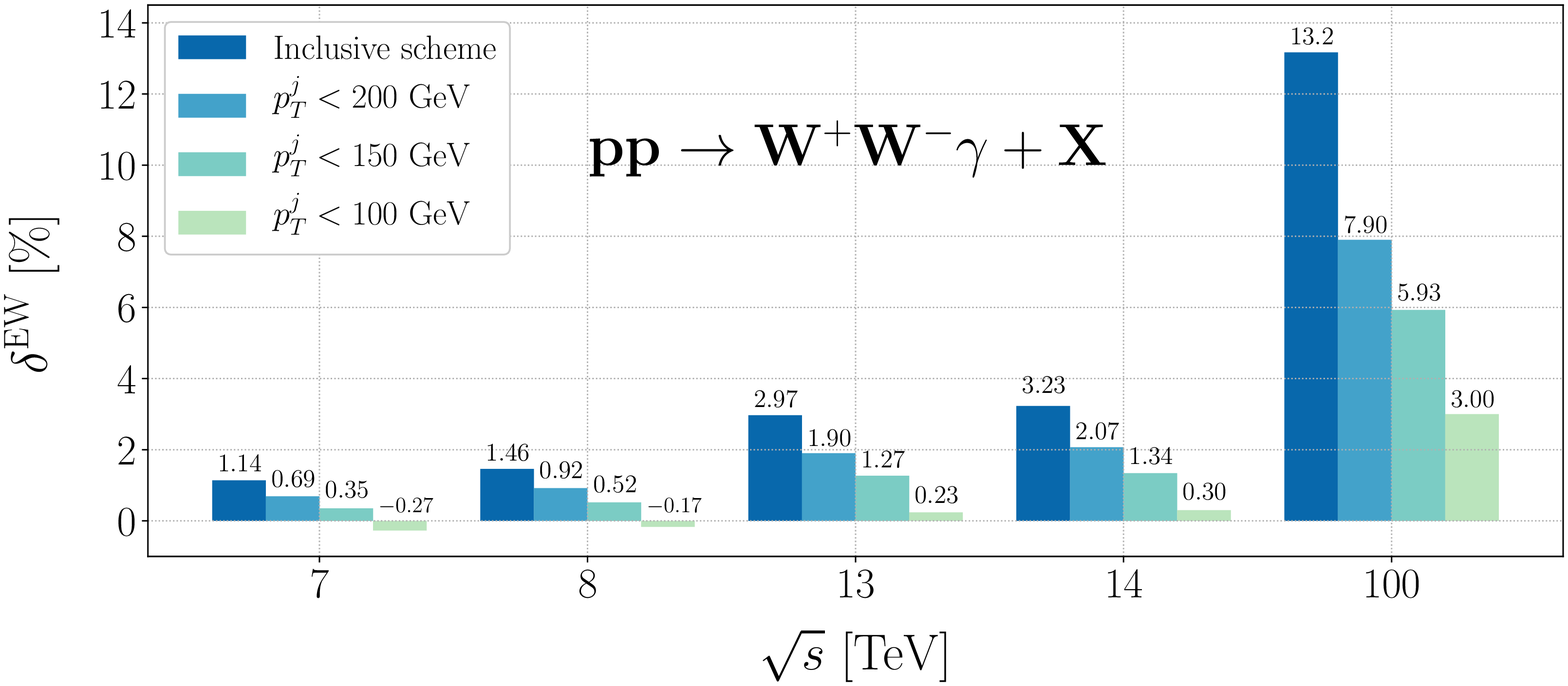}
\caption{NLO QCD (upper panel) and EW (lower panel) relative corrections to $pp \rightarrow W^+W^-\gamma + X$ in both inclusive and exclusive event selection schemes for some typical values of $p_T^{j,\, {\rm cut}}$ and $\sqrt{s}$.}
\centering
\label{relative-ex}
\end{figure}

Due to the large contribution from the real jet emission channels (listed in the right side of Table \ref{Real-emission}), the QCD and EW relative corrections can be obviously suppressed by applying an additional transverse momentum cut on the final-state jet, i.e., $p_{T}^{j} < p_{T}^{j,\, {\rm cut}}$. We call this suppression scheme as the exclusive event selection scheme. In the exclusive event selection scheme, the cancellation of IR singularities is still held since the soft region of emitted jet is remained. In Table \ref{EX-CS} we present the NLO QCD+EW corrected integrated cross sections in both exclusive and inclusive event selection schemes for comparison, where the jet transverse momentum cut (i.e., the upper bound of the jet transverse momentum) is taken as $p_T^{j,\, {\rm cut}} =100~ {\rm GeV}$, $150~ {\rm GeV}$, $200~ {\rm GeV}$ and $+\infty$, separately\footnote{$p_T^{j,\, {\rm cut}} = +\infty$ corresponds to the inclusive event selection scheme.}. The corresponding QCD and EW relative corrections are provided in the upper and lower plots of Fig.\ref{relative-ex}, respectively. The two plots clearly demonstrate that the NLO QCD and EW relative corrections, $\delta^{{\rm QCD}}$ and $\delta^{{\rm EW}}$, increase with the increment of the $pp$ colliding energy, and decrease with the decline of the jet transverse momentum cut. At the $7~ {\rm TeV}$ LHC, the QCD and EW relative corrections in the inclusive event selection scheme are $61.5\%$ and $1.14\%$, respectively, and can be reduced to $42.3\%$ and $-0.27\%$ after applying $p_{T}^{j} < 100~ {\rm GeV}$ on the final-state jet. While at the $14~ {\rm TeV}$ LHC, the QCD and EW relative corrections decrease from $84.0\%$ and $3.23\%$ in the inclusive event selection scheme to $47.3\%$ and $0.30\%$ in the $p_T^{j,\, {\rm cut}} = 100~ {\rm GeV}$ exclusive event selection scheme, respectively. The full EW relative correction at current LHC is very small and could be neglected to some extent in the $p_T^{j,\, {\rm cut}} = 100~ {\rm GeV}$ exclusive event selection scheme in experimental analysis. However, when the colliding energy is raised to $100~ {\rm TeV}$, we get $\delta^{\text{EW}}=3.00\%$ in the $p_T^{j,\, {\rm cut}}=100~ {\rm GeV}$ exclusive event selection scheme, and thus the EW correction can not be ignored in precise experiments.
\begin{table}[htbp]
\begin{center}
\renewcommand\arraystretch{1.8}
\begin{tabular}{ccccc}
\hline
\hline
\multirow{2}*{$\sqrt{s}$ [TeV]}
& \multicolumn{3}{c}{Exclusive scheme $\sigma^{{\rm NLO}}$ [pb]} & \multirow{2}*{Inclusive scheme $\sigma^{{\rm NLO}}$ [pb]} \\
\cline{2-4}
& $p_T^{j,\, {\rm cut}} = 100~ {\rm GeV}$ & $p_T^{j,\, {\rm cut}} = 150~ {\rm GeV}$ & $p_T^{j,\, {\rm cut}} = 200~ {\rm GeV}$ & \\
\hline
7  & 0.0942(1) & 0.1002(1) & 0.1035(1) & 0.1074(1) \\
8  & 0.1165(1) & 0.1249(1) & 0.1295(1) & 0.1356(1) \\
13 & 0.2385(3) & 0.2611(3) & 0.2738(3) & 0.2971(3) \\
14 & 0.2640(3) & 0.2901(3) & 0.3041(3) & 0.3327(4) \\
100& 2.573(3)  & 3.027(3)  & 3.322(3)  & 4.237(4)  \\
\hline
\hline
\end{tabular}
\caption{NLO QCD+EW corrected cross sections for $pp \rightarrow W^+W^-\gamma + X$ in both inclusive and exclusive event selection schemes at different $pp$ colliding energies.}
\label{EX-CS}
\end{center}
\end{table}

\subsection{Kinematic distributions}
\par
Now we turn to the kinematic distributions of the final-state $W^{\pm}$ bosons for $pp \rightarrow W^+W^-\gamma + X$ at the $14~ {\rm TeV}$ LHC in the inclusive event selection scheme.

\par
\begin{figure}[htbp]
\centering
\includegraphics[scale=0.4]{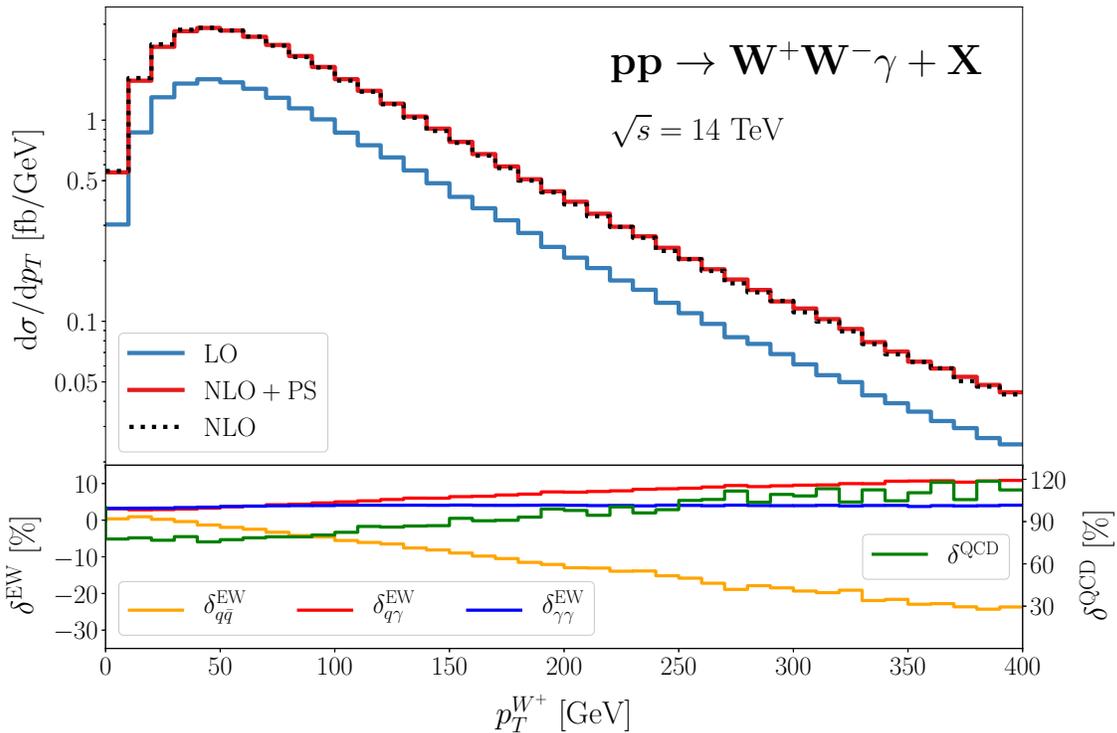}
\caption{
(upper panel) LO, NLO and shower-matched NLO corrected transverse momentum distributions of final $W^+$ boson for $W^+W^-\gamma$ production at the $14~ {\rm TeV}$ LHC in the inclusive event selection scheme.
(lower panel) Corresponding QCD and $q\bar{q}$-, $q\gamma$-, $\gamma\gamma$-initiated EW relative corrections.}
\centering
\label{PT}
\end{figure}

The LO, NLO and shower-matched NLO (NLO+PS) corrected transverse momentum distributions of $W^+$ boson are plotted in the upper panel of Fig.\ref{PT}. The corresponding EW relative corrections from the $q\bar{q}$-, $q\gamma$- and $\gamma\gamma$-initiated channels as well as the QCD relative correction are provided in the lower panel of Fig.\ref{PT}. The shower-matched NLO corrected cross section is calculated by
\begin{eqnarray}
\sigma^{{\rm NLO+PS}}
=
\sigma^{{\rm LO}}
\times
\Big[
\left(1 + \delta_{q\bar{q}}^{{\rm EW}}\right)
\left(1 + \delta^{{\rm QCD+PS}}\right)
+
\delta_{q\gamma}^{{\rm EW}}
+
\delta_{\gamma\gamma}^{{\rm EW}}
\Big],
\end{eqnarray}
where $\delta^{{\rm QCD+PS}}$ is shower-matched NLO QCD relative correction. The LO and (shower-matched) NLO QCD+EW corrected $p_T^{W^+}$ distributions increase rapidly in the low $p_T^{W^+}$ region, reach their maxima in the vicinity of $p_T^{W^+} \sim 45~ {\rm GeV}$, and decrease approximately logarithmically when $p_T^{W^+} > 60~ {\rm GeV}$ as the increment of $p_T^{W^+}$. The NLO QCD correction significantly enhances the LO $W^+$-boson transverse momentum distribution in the whole plotted $p_T^{W^+}$ region. The corresponding QCD relative correction increases from about $77\%$ to approximately $120\%$ as the increment of $p_T^{W^+}$ from $0$ to $400~ {\rm GeV}$. The $q\gamma$-induced relative correction is positively correlated with $p_T^{W^+}$ and can exceed $10\%$ when $p_T^{W^+} > 350~ {\rm GeV}$, while the $\gamma\gamma$-induced relative correction is remarkably stable as $p_T^{W^+}$ varies in the region of $p_T^{W^+} < 400~ {\rm GeV}$. The EW relative correction from the $q\bar{q}$-initiated channel decreases consistently as the increment of $p_T^{W^+}$ and reaches about $-20\%$ at $p_T^{W^+} = 300~ {\rm GeV}$ due to the large EW Sudakov logarithms induced by the virtual massive gauge bosons in loops. It is clear that the EW correction from the $q\bar{q}$ annihilation channel can be almost compensated by the positive photon-induced (i.e., $q\gamma$- and $\gamma\gamma$-induced) corrections in the high $p_T^{W^+}$ region. The ``average'' PS relative correction to a NLO corrected kinematic distribution of observable ${\cal O}$ in the region of $a \leqslant {\cal O} \leqslant b$ is defined as
\begin{eqnarray}
\delta^{\text{PS}}_{{\cal O} \in [a,\, b]}
= 100\%
\times
\int_a^b
\left(
\frac{{\rm d} \sigma^{{\rm NLO+PS}}}{{\rm d} {\cal O}}
-
\frac{{\rm d}\sigma^{{\rm NLO}}}{{\rm d} {\cal O}}
\right)
{\rm d} {\cal O}
\bigg/
\int_a^b
\frac{{\rm d}\sigma^{{\rm NLO}}}{{\rm d} {\cal O}}
{\rm d} {\cal O}.
\end{eqnarray}
From the figure we can see that the PS correction slightly suppresses and enhances the $p_T^{W^+}$ distribution in the low and high $p_T^{W^+}$ regions, respectively. In the region of $p_T^{W^+} \in [0,\, 50]~ {\rm GeV}$, the average PS relative correction to the NLO QCD+EW corrected $p_T^{W^+}$ distribution is $\delta^{{\rm PS}}_{p_T^{W^+} \in\, [0,\, 50]\,{\rm GeV}} \simeq -1.7\%$. This PS relative correction is the same order as the EW relative correction, and should be taken into consideration in precision predictions.

\par
\begin{figure}[htbp]
\centering
\includegraphics[scale=0.4]{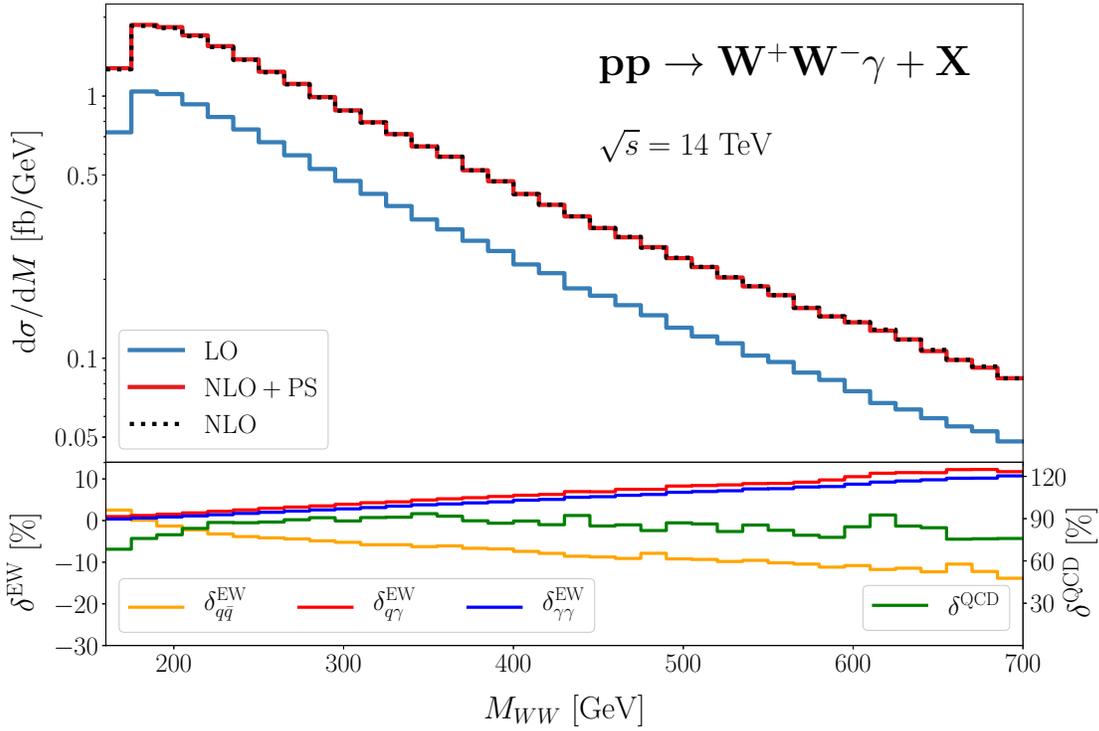}
\caption{The same as Fig.\ref{PT}, but for the invariant mass distribution of $W$-boson pair.}
\centering
\label{M}
\end{figure}

The LO, NLO and NLO+PS corrected invariant mass distributions of the final $W$-boson pair are presented in the upper panel, and the corresponding EW relative corrections from various partonic channels as well as the QCD relative correction are depicted in the lower panel of Fig.\ref{M}. In the plotted $M_{WW}$ region, the PS effect on the $W^+W^-$ invariant mass distribution is very small, and the QCD relative correction varies in the range of $[70\%,\, 90\%]$ approximately. The EW relative corrections from the $q\bar{q}$-, $q\gamma$- and $\gamma\gamma$-initiated channels are strongly dependent on the invariant mass of the $W^+W^-$ system. As the increment of $M_{WW}$ from its threshold to $700~ {\rm GeV}$, the $\gamma\gamma$- and $q\gamma$-induced relative corrections increase approximately linearly from about $0$ and $1\%$ to about $11\%$ and $12\%$, respectively. By contrast, the EW relative correction from the $q\bar{q}$ annihilation channel decreases with the increment of $M_{WW}$. It exceeds $-10\%$ when $M_{WW} > 550~ {\rm GeV}$ and can be close to $-15\%$ at $M_{WW} = 700~ {\rm GeV}$ due to the large Sudakov EW logarithms. As we expected, the negative EW correction from the $q\bar{q}$ annihilation channel can be canceled out by the positive photon-induced corrections in the high invariant mass region.

\par
\begin{figure}[htbp]
\centering
\includegraphics[scale=0.4]{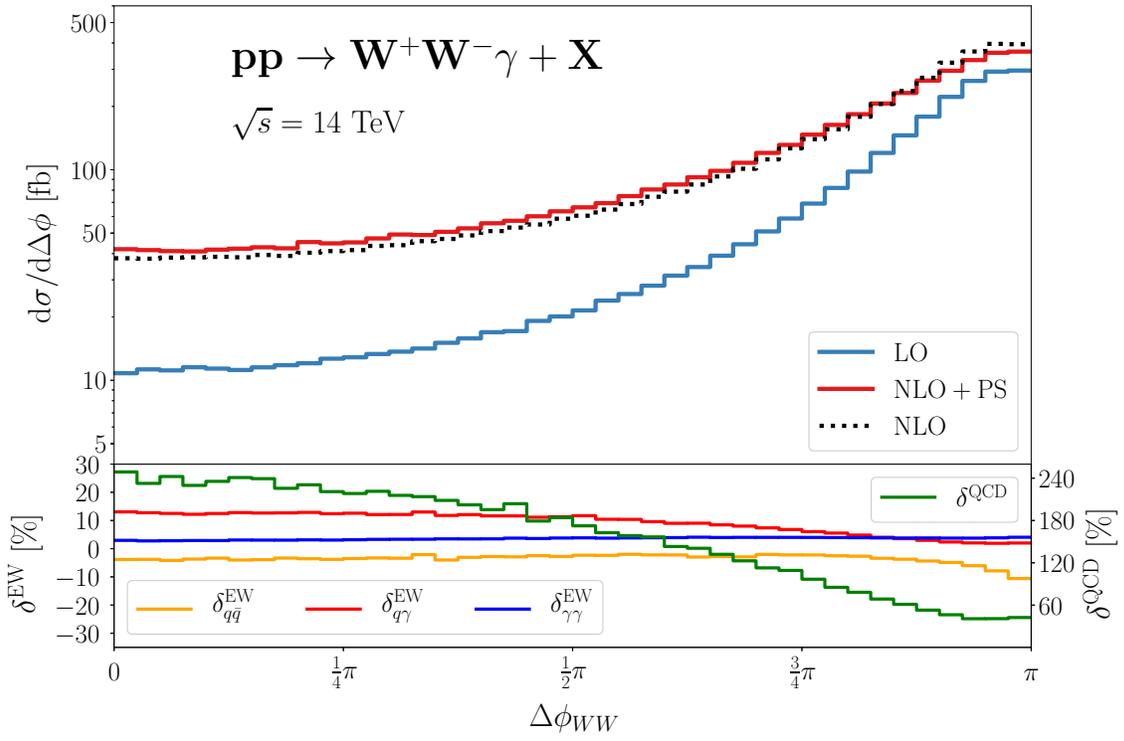}
\caption{The same as Fig.\ref{PT}, but for the distribution of the azimuthal angle difference between the two final $W$ bosons.}
\centering
\label{dphi}
\end{figure}

The LO, NLO and NLO+PS corrected distributions of the azimuthal angle difference between the two final $W$ bosons ($\Delta \phi_{WW}$) are depicted in the upper panel of Fig.\ref{dphi}, and the corresponding QCD and EW relative corrections ($\delta^{{\rm QCD}}$, $\delta_{q\bar{q}}^{{\rm EW}}$, $\delta_{q\gamma}^{{\rm EW}}$ and $\delta_{\gamma\gamma}^{{\rm EW}}$) are plotted in the lower panel of Fig.\ref{dphi}. Both the LO and (shower-matched) NLO corrected $\Delta \phi_{WW}$ distributions increase consistently as the increment of $\Delta \phi_{WW}$, and the $W^+W^-\gamma$ events are more concentrated in the vicinity of $\Delta \phi_{WW} \sim \pi$. That means the two final-state $W$ bosons prefer to be produced back-to-back in the transverse plane. Compared to the $p_T^{W^+}$ and $M_{WW}$ distributions, the PS effect on the $\Delta \phi_{WW}$ distribution is more remarkable. The PS correction enhances and suppresses the $\Delta \phi_{WW}$ distribution in the regions of $\Delta \phi_{WW} < 0.85 \pi$ and $\Delta \phi_{WW} > 0.85 \pi$, respectively. In the region of $\Delta \phi_{WW} \in [0.9\pi,\, \pi]$, the average PS relative correction to the NLO QCD+EW corrected $\Delta \phi_{WW}$ distribution is $\delta^{{\rm PS}}_{\Delta\phi_{WW} \in\, [0.9\pi,\, \pi]} \simeq -8.6\%$, which is a more considerable correction factor compared to the full EW relative correction in this specific phase-space region. The NLO QCD correction enhances the LO $\Delta \phi_{WW}$ distribution significantly, especially in the region with small azimuthal angle difference. The corresponding QCD relative correction approximately decreases from $250\%$ to $40\%$ as the increment of $\Delta\phi_{WW}$ from $0$ to $\pi$. The $q\gamma$-induced relative correction holds steady at $11\% \sim 13\%$ in the region of $\Delta\phi_{WW} < \pi/2$ and then gradually decreases to about $2\%$ as $\Delta\phi_{WW}$ increases to $\pi$, while the $\gamma\gamma$-induced relative correction is steady at $3\% \sim 4\%$ in the entire $\Delta\phi_{WW}$ region. Different from the photon-induced relative corrections, the EW relative correction from the $q\bar{q}$ annihilation channel is negative. It can exceed $-10\%$ when the two final $W$ bosons are sufficiently anticollinear in the transverse plane (i.e., $\Delta\phi_{WW} \rightarrow \pi$).

\subsection{Theoretical uncertainty}
\begin{figure}[htbp]
\centering
\includegraphics[scale=0.4]{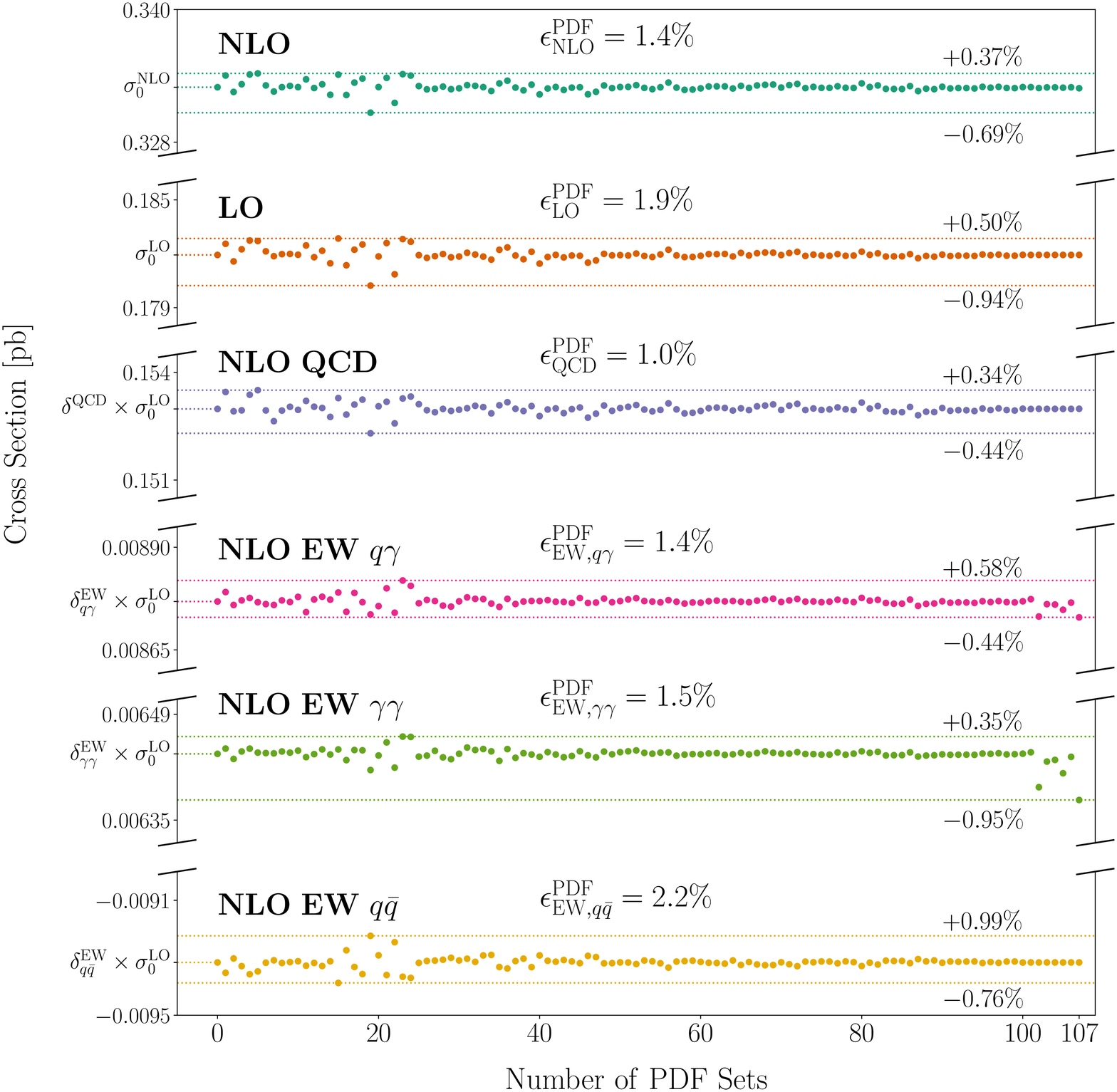}
\caption{
LO and NLO QCD+EW corrected cross sections, as well as the QCD and $q\bar{q}$-, $q\gamma$- and $\gamma\gamma$-initiated EW corrections, for $pp \rightarrow W^+W^-\gamma + X$ at the $14~ {\rm TeV}$ LHC in the inclusive event selection scheme obtained by using central and eigenvector sets of the {\sc LUXqed} PDF.}
\centering
\label{pdf}
\end{figure}

The uncertainty of PDFs is a main source of theoretical error for scattering processes at hadron colliders. In this work, We use the LUXqed\_plus\_PDF4LHC15\_nnlo\_100 PDF to perform the initial-state parton convolution. {\sc LUXqed} is a Hessian PDF, it contains a central PDF set and $N = 107$ eigenvector PDF sets (i.e., error PDF sets). The PDF uncertainty of a cross section $\sigma$ calculated with the {\sc LUXqed} PDF is given by
\begin{eqnarray}
\label{pdf-def-Hessian}
\epsilon^{{\rm PDF}}
=
\dfrac{1}{\sigma_0}
\sqrt{\sum_{j=1}^N (\sigma_j - \sigma_0)^2},
\end{eqnarray}
where $\sigma_j~ (j = 1, ..., N)$ is the cross section evaluated with eigenvector set $j$ and $\sigma_{0}$ is the central value calculated with central set. To demonstrate the theoretical errors induced by the {\sc LUXqed} PDF more clearly, in Fig.\ref{pdf}, we depict the LO and NLO QCD+EW corrected cross sections, as well as the QCD and $q\bar{q}$-, $q\gamma$- and $\gamma\gamma$-initiated EW corrections, for $pp \rightarrow W^+W^-\gamma + X$ at the $14~ {\rm TeV}$ LHC in the inclusive event selection scheme obtained with each {\sc LUXqed} PDF set. The figure shows that the PDF uncertainty of the inclusive cross section for $pp \rightarrow W^+W^-\gamma + X$ at the $14~ {\rm TeV}$ LHC is $1.9\%$ at the LO, and can be reduced to $1.4\%$ if the NLO QCD and EW corrections are taken into account. The PDF uncertainty of the NLO QCD correction is much smaller than that of the LO cross section, it is only about half of the PDF uncertainty of the LO cross section ($\epsilon^{{\rm PDF}}_{{\rm QCD}} = 1.0\% \simeq \dfrac{1}{2} \times \epsilon^{{\rm PDF}}_{{\rm LO}}$). As for the NLO EW correction, the PDF uncertainties of the photon-induced and $q\bar{q}$-initiated contributions  are slightly smaller and larger than that of the LO cross section, respectively ($\epsilon^{{\rm PDF}}_{{\rm EW}, q\gamma} = 1.4\%$, $\epsilon^{{\rm PDF}}_{{\rm EW}, \gamma\gamma} = 1.5\%$ and $\epsilon^{{\rm PDF}}_{q\bar{q}} = 2.2\%$). Compared to the $q\bar{q}$-, $q\gamma$- and $\gamma\gamma$-initiated EW corrections, the QCD correction is the dominant contribution at the NLO, and its PDF uncertainty is the smallest. Thus, the PDF uncertainty of the LO cross section can be reduced by the NLO correction, even though the relative error of the $q\bar{q}$-initiated EW correction induced by the {\sc LUXqed} PDF is larger than that of the LO cross section. This improvement of the PDF uncertainty at the NLO is mainly due to the QCD correction. Moreover, it should be noted that the LO cross section, NLO QCD correction and $q\bar{q}$-initiated EW correction are independent of the last seven ($j = 101, ..., 107$) eigenvector PDF sets. That is because the last seven eigenvector sets of the {\sc LUXqed} PDF provide the same distribution functions for QCD partons.

\par
To assess the theoretical error from the uncertainty of the photon distribution function, we analyze the photon-induced channels (i.e., $q\gamma$- and $\gamma\gamma$-initiated channels) with the following PDF sets incorporating QED corrections for comparison:
\begin{itemize}
\item
\underline{MRST2004qed\_proton ({\sc MRST2004qed})} \\
The {\sc MRST2004qed} \cite{MRST} PDF does not provide any information on the PDF uncertainty, thus we do not use it in the uncertainty estimate.
\item
\underline{CT14qed\_inc\_proton ({\sc CT14qed})} \\
The {\sc CT14qed} \cite{CT14} PDF contains $31$ PDF sets: CT14qed\_inc\_proton\_{\it id}.dat ({\it id} = 0000, ..., 0030). The central value is calculated by the central set CT14qed\_inc\_proton\_0000.dat. The photon-induced cross section is monotonically dependent on the {\it id} of the {\sc CT14qed} PDF set, thus the theoretical error induced by this PDF can be quantified by the difference of the two results obtained with the CT14qed\_inc\_proton\_0000.dat and CT14qed\_inc\_proton\_0030.dat PDF sets, respectively.
\item
\underline{NNPDF31\_nnlo\_as\_0118\_luxqed ({\sc NNPDF31})} \\
{\sc NNPDF31} \cite{NNPDF} is a Monte Carlo PDF, it provides $N_{{\rm rep}} = 100$ Monte Carlo replicas of PDFs. The PDF uncertainty of a cross section $\sigma$ evaluated with the {\sc NNPDF31} PDF is given by
\begin{eqnarray}
\label{pdf-def-MC}
\epsilon^{{\rm PDF}}
=
\dfrac{1}{\sigma_0}
\sqrt{\dfrac{1}{N_{{\rm rep}} - 1} \sum_{j=1}^{N_{{\rm rep}}} (\sigma_j - \sigma_0)^2},
\end{eqnarray}
where $\sigma_j~ (j = 1, ..., N_{{\rm rep}})$ is the cross section evaluated with replica $j$ and $\sigma_0$ the central value of the cross section.
\item
\underline{LUXqed\_plus\_PDF4LHC15\_nnlo\_100 ({\sc LUXqed})} \\
{\sc LUXqed} is a Hessian PDF. The PDF uncertainty of a cross section is given by Eq.(\ref{pdf-def-Hessian}).
\end{itemize}
We list the $q\gamma$- and $\gamma\gamma$-induced relative corrections and the corresponding PDF uncertainties for the inclusive production of $W^+W^-\gamma$ at the $14~ {\rm TeV}$ LHC obtained with the above PDF sets separately in Table \ref{PDFs}. For the {\sc CT14qed} PDF, we provide both the central values and the results obtained with the CT14qed\_inc\_proton\_0030.dat PDF set. We can see that the largely outdated {\sc MRST2004qed} PDF obviously overestimates the photon-induced relative corrections compared to the {\sc NNPDF31} and {\sc LUXqed} PDFs, and moreover, it can not provide the theoretical errors from the uncertainty of PDFs. For $q\gamma$- and $\gamma\gamma$-initiated channels, the PDF errors induced by {\sc CT14qed} are significantly larger than the corresponding ones induced by {\sc LUXqed} and {\sc NNPDF31}. That is because the {\sc CT14qed} PDF has a LO evolution of the photon and uses ZEUS data for the fit, while the {\sc LUXqed} and {\sc NNPDF31} PDFs are basically the same set, with some NLO photon fitting, and different data sets for the photon than {\sc CT14qed}. For the {\sc NNPDF31} PDF, the photon content in the proton is supplemented by {\sc LUX} in its fitting procedure. Thus, as we expected, the central values for the $q\gamma$- and $\gamma\gamma$-induced relative corrections obtained from the {\sc NNPDF31} PDF are almost the same as the corresponding ones from the {\sc LUXqed} PDF. Moreover, the PDF uncertainties of the photon-induced corrections given by the {\sc NNPDF31} and {\sc LUXqed} PDFs are only about $1\% \sim 1.5\%$. The PDF errors in the {\sc NNPDF31} and {\sc LUXqed} PDFs are almost certainly understated, but they are in principle NLO fits, and so they should be used instead of the {\sc CT14qed} PDF because they match the correct order of our process. Thus, the {\sc LUXqed} PDF as well as the {\sc NNPDF31} PDF provides an optimal choice for precision study on the $W^+W^-\gamma$ production via the $q\gamma$ scattering and $\gamma\gamma$ fusion channels at the LHC.
\begin{table}[htbp]
\begin{center}
\renewcommand\arraystretch{1.8}
\begin{tabular}{ccccc}
\hline
\hline
PDF set & {\sc MRST2004qed} & {\sc CT14qed} & {\sc NNPDF31} & {\sc LUXqed} \\
\hline
$\delta_{q\gamma}^{{\rm EW}}$ [\%]       & $6.01$ & $4.23-7.46$ & $4.81^{+0.05}_{-0.05}$ & $4.81^{+0.07}_{-0.07}$ \\
$\delta_{\gamma\gamma}^{{\rm EW}}$ [\%]  & $4.95$ & $2.79-8.28$ & $3.51^{+0.05}_{-0.05}$ & $3.54^{+0.05}_{-0.05}$ \\
\hline
\hline
\end{tabular}
\caption{
$q\gamma$- and $\gamma\gamma$-induced relative corrections and the corresponding PDF uncertainties for the inclusive production of $W^+W^-\gamma$ at the $14~ {\rm TeV}$ LHC obtained with {\sc MRST2004qed}, {\sc CT14qed}, {\sc NNPDF31} and {\sc LUXqed}, separately.}
\label{PDFs}
\end{center}
\end{table}

\par
Another important source of theoretical error for scattering processes at hadron colliders is the factorization/renormalization scale dependence. For the $pp \rightarrow W^+W^-\gamma + X$ process, the factorization scale $\mu_f$ is involved in all perturbative orders via the initial-state parton convolution, while the renormalization scale $\mu_r$ appears only at high orders since the $W^+W^-\gamma$ production at the LO is an EW process. In Table \ref{scale}, we present the numerical results of the LO and NLO QCD+EW corrected cross sections for $W^+W^-\gamma$ production at the $14~ {\rm TeV}$ LHC in the inclusive event selection scheme for some typical value of $\kappa_f$ and $\kappa_r$, where $\kappa_{f} \equiv \mu_{f}/\mu_0$ and $\kappa_{r} \equiv \mu_{r}/\mu_0$. In this work, the scale uncertainty of an integrated cross section is defined as
\begin{eqnarray}
\epsilon^{{\rm scale}}
=
\frac{1}{\sigma(\mu_0, \mu_0)}
\times
\max
\left\{
\left[ \sigma(\mu_f, \mu_r) - \sigma(\mu_f^{\prime}, \mu_r^{\prime}) \right]
\Big|~ \mu_f,\, \mu_f^{\prime},\, \mu_r,\, \mu_r^{\prime} \in \big\{ \mu_0/2,\, \mu_0,\, 2\mu_0 \big\}
\right\}.
\end{eqnarray}
Then, from Table \ref{scale}, we obtain the scale uncertainties of the LO and NLO QCD+EW corrected inclusive integrated cross sections for $pp \rightarrow W^+W^-\gamma + X$ at the $14~ {\rm TeV}$ LHC as
\begin{eqnarray}
\epsilon^{{\rm scale}}_{{\rm LO}} = 5.0\%, ~~~~~~~~~~~~~~~ \epsilon^{{\rm scale}}_{{\rm NLO}} = 14.5\%.
\end{eqnarray}
We can see that the scale uncertainty at the LO is only about one third of that at the QCD+EW NLO for the inclusive production of $W^+W^-\gamma$ at the $14~ {\rm TeV}$ LHC. Since the strong interaction is not involved in the $W^+W^-\gamma$ production at the LO, the LO scale uncertainty would underestimate the theoretical error due to missing higher order contributions. The NLO QCD+EW corrected cross section is more sensitive to the renormalization scale than the LO cross section, and the scale sensitivity can be depressed by including higher order radiative corrections. Finally, we see that the scale uncertainty is the main source of theoretical error. It is about one order of magnitude larger than the PDF uncertainty at the NLO.
\begin{table}[htbp]
\begin{center}
\renewcommand\arraystretch{1.8}
\begin{tabular}{ccccc}
\hline
\hline
\multirow{2}*{$\kappa_f$}
& \multicolumn{3}{c}{$\sigma^{{\rm NLO}}$ [pb]} & \multirow{2}*{$\sigma^{{\rm LO}}$ [pb]} \\
\cline{2-4}
& $\kappa_r = 1/2$ & $\kappa_r = 1$ & $\kappa_r = 2$ & \\
\hline
~~1/2~~ & ~~0.3474(4)~~ & ~~0.3529(4)~~ & ~~0.3582(4)~~ & ~~0.17659(5)~~ \\
1   & 0.3271(4) & 0.3327(4) & 0.3385(4) & 0.18190(5) \\
2   & 0.3100(4) & 0.3160(4) & 0.3217(4) & 0.18575(5) \\
\hline
\hline
\end{tabular}
\caption{Factorization/renormalization scale dependence of the LO and NLO QCD+EW corrected cross sections for $pp \rightarrow W^+W^-\gamma + X$ at the $14~ {\rm TeV}$ LHC in the inclusive event selection scheme.}
\label{scale}
\end{center}
\end{table}

\vskip 10mm
\section{Summary}
\par
In this work, we calculate the NLO QCD and EW corrections to $pp \rightarrow W^+W^-\gamma + X$, and combine these corrections by using the multiplying approximation in Eq.(\ref{NLO-exp}) to obtain the full NLO QCD+EW corrected theoretical predictions at some specific colliding energies. At the NLO, the QCD correction is the dominant contribution, however, the EW correction is also considerable at current LHC and future proton-proton colliders. The large QCD and EW corrections from the real jet emission channels, $q\bar{q} \rightarrow W^+W^-\gamma g$, $qg \rightarrow W^+W^-\gamma q$ and $q\gamma \rightarrow W^+W^-\gamma q$, can be depressed sufficiently by applying the jet veto event selection scheme. At the LHC, the positive EW correction from the $q\gamma$- and $\gamma\gamma$-initiated channels and the negative EW correction from the $q\bar{q}$ annihilation channel almost cancel each other out, and the residual NLO EW relative correction is less than $1\%$, if the transverse momentum cut of $p_T^{j} < 100~ {\rm GeV}$ is applied on the final-state jet. We also provide some kinematic distributions at the QCD+EW NLO including parton shower effects. The Sudakov effects in $W^+W^-\gamma$ production at the LHC are clearly shown at high $p_T^{W^+}$ and $M_{WW}$ regions, and the parton shower effects are observable in the distribution of the azimuthal angle difference between the two final $W$ bosons. Moreover, we present a detailed investigation on the theoretical errors arising from the PDF uncertainty and the factorization/renormalization scale dependence. We find that the {\sc LUXqed} PDF as well as the {\sc NNPDF31} PDF is more suitable for precision study of the $q\gamma$- and $\gamma\gamma$-initiated channels, and the theoretical error of the NLO corrected cross section is dominated by the factorization/renormalization scale uncertainty.

\vskip 10mm
\par
\noindent{\large\bf ACKNOWLEDGMENTS} \\
This work is supported in part by the National Natural Science Foundation of China (Grants No. 11775211 and No. 11535002) and the CAS Center for Excellence in Particle Physics (CCEPP).

\end{document}